\PassOptionsToPackage{unicode}{hyperref}
\PassOptionsToPackage{hyphens}{url}
\PassOptionsToPackage{dvipsnames,svgnames,x11names}{xcolor}
\documentclass[12pt]{article}

\usepackage{multirow}
\usepackage{multicol}

\usepackage{graphics}
\usepackage{graphicx,psfrag,epsf}
\usepackage{enumerate}

\usepackage{amssymb}
\usepackage{bm}
\usepackage{color}
\usepackage{amsthm}
\usepackage{amsfonts}
\usepackage{subfig}
\usepackage{float}
\usepackage{comment}
\usepackage{ulem}
\usepackage{longtable}
\usepackage{tabularx}
\usepackage{array}

\usepackage{amsmath,amssymb}
\usepackage{iftex}

\newcommand{\shiftedlongtable}[2]{%
  {% Begin local group
  \setlength{\LTleft}{#1}%
  #2%
  }% End local group
}

\ifPDFTeX
  \usepackage[T1]{fontenc}
  \usepackage[utf8]{inputenc}
  \usepackage{textcomp} % provide euro and other symbols
\else % if luatex or xetex
  \usepackage{unicode-math}
  \defaultfontfeatures{Scale=MatchLowercase}
  \defaultfontfeatures[\rmfamily]{Ligatures=TeX,Scale=1}
\fi
\usepackage{lmodern}
\ifPDFTeX\else  
\fi
\IfFileExists{upquote.sty}{\usepackage{upquote}}{}
\IfFileExists{microtype.sty}{% use microtype if available
  \usepackage[]{microtype}
  \UseMicrotypeSet[protrusion]{basicmath} % disable protrusion for tt fonts
}{}
\makeatletter
\@ifundefined{KOMAClassName}{% if non-KOMA class
  \IfFileExists{parskip.sty}{%
    \usepackage{parskip}
  }{% else
    \setlength{\parindent}{0pt}
    \setlength{\parskip}{6pt plus 2pt minus 1pt}}
}{% if KOMA class
  \KOMAoptions{parskip=half}}
\makeatother
\usepackage{xcolor}
\makeatletter
\ifx\paragraph\undefined\else
  \let\oldparagraph\paragraph
  \renewcommand{\paragraph}{
    \@ifstar
      \xxxParagraphStar
      \xxxParagraphNoStar
  }
  \newcommand{\xxxParagraphStar}[1]{\oldparagraph*{#1}\mbox{}}
  \newcommand{\xxxParagraphNoStar}[1]{\oldparagraph{#1}\mbox{}}
\fi
\ifx\subparagraph\undefined\else
  \let\oldsubparagraph\subparagraph
  \renewcommand{\subparagraph}{
    \@ifstar
      \xxxSubParagraphStar
      \xxxSubParagraphNoStar
  }
  \newcommand{\xxxSubParagraphStar}[1]{\oldsubparagraph*{#1}\mbox{}}
  \newcommand{\xxxSubParagraphNoStar}[1]{\oldsubparagraph{#1}\mbox{}}
\fi
\makeatother

\usepackage{longtable,booktabs,array}
\usepackage{calc} % for calculating minipage widths
\usepackage{etoolbox}
\makeatletter
\patchcmd\longtable{\par}{\if@noskipsec\mbox{}\fi\par}{}{}
\makeatother
\IfFileExists{footnotehyper.sty}{\usepackage{footnotehyper}}{\usepackage{footnote}}
\makesavenoteenv{longtable}
\usepackage{graphicx}
\makeatletter
\def\maxwidth{\ifdim\Gin@nat@width>\linewidth\linewidth\else\Gin@nat@width\fi}
\def\maxheight{\ifdim\Gin@nat@height>\textheight\textheight\else\Gin@nat@height\fi}
\makeatother
\setkeys{Gin}{width=\maxwidth,height=\maxheight,keepaspectratio}
\makeatletter
\def\fps@figure{htbp}
\makeatother

\makeatletter
\@ifpackageloaded{caption}{}{\usepackage{caption}}
\AtBeginDocument{%
\ifdefined\contentsname
  \renewcommand*\contentsname{Table of contents}
\else
  \newcommand\contentsname{Table of contents}
\fi
\ifdefined\listfigurename
  \renewcommand*\listfigurename{List of Figures}
\else
  \newcommand\listfigurename{List of Figures}
\fi
\ifdefined\listtablename
  \renewcommand*\listtablename{List of Tables}
\else
  \newcommand\listtablename{List of Tables}
\fi
\ifdefined\figurename
  \renewcommand*\figurename{Figure}
\else
  \newcommand\figurename{Figure}
\fi
\ifdefined\tablename
  \renewcommand*\tablename{Table}
\else
  \newcommand\tablename{Table}
\fi
}
\@ifpackageloaded{float}{}{\usepackage{float}}
\floatstyle{ruled}
\@ifundefined{c@chapter}{\newfloat{codelisting}{h}{lop}}{\newfloat{codelisting}{h}{lop}[chapter]}
\floatname{codelisting}{Listing}

\makeatother
\makeatletter
\@ifpackageloaded{caption}{}{\usepackage{caption}}
\@ifpackageloaded{subcaption}{}{\usepackage{subcaption}}
\makeatother

\ifLuaTeX
  \usepackage{selnolig}  % disable illegal ligatures
\fi
\usepackage[]{natbib}
\usepackage{bookmark}

\IfFileExists{xurl.sty}{\usepackage{xurl}}{} % add URL line breaks if available
\hypersetup{
  pdftitle={Title},
  pdfauthor={Author 1; Author 2},
  pdfkeywords={3 to 6 keywords, that do not appear in the title},
  colorlinks=true,
  linkcolor={blue},
  filecolor={Maroon},
  citecolor={Blue},
  urlcolor={Blue},
  pdfcreator={LaTeX via pandoc}}

\newcommand{\anon}{1}

\begin{document}

\def\spacingset#1{\renewcommand{\baselinestretch}%
{#1}\small\normalsize} \spacingset{1}

%%%%%%%%%%%%%%%%%%%%%%%%%%%%%%%%%%%%%%%%%%%%%%%%%%%%%%%%%%%%%%%%%%%%%%%%%%%%%%

\if1\anon
{
  \title{\bf Estimating Spatially-Smoothed Fiber Orientation Distribution from Diffusion-MRI Experiments}
  \author{Jilei Yang \\
    Department of Statistics, University of California,  Davis\\
    Seungyong Hwang\thanks{Corresponding author: syhwang@jbnu.ac.kr}\\
    Department of Statistics and Institute of Applied Statistics,\\
    Jeonbuk National University\\
    Mengjie
    Shi\\
    Department of Statistics, University of California,  Davis\\
    and \\
    Jie Peng\\
    Department of Statistics, University of California,  Davis}
  \maketitle
} \fi

\if0\anon
{
  \bigskip
  \bigskip
  \bigskip
  \begin{center}
    {\LARGE\bf Estimating Spatially-Smoothed Fiber Orientation Distribution from Diffusion-MRI Experiments}
\end{center}
  \medskip
} \fi

\bigskip
\begin{abstract}
Diffusion-weighted magnetic resonance imaging (D-MRI) is a noninvasive in vivo technique for probing the microstructural architecture of biological tissues. At each voxel, the fiber orientation distribution (FOD) characterizes local fiber configurations and orientations and is therefore a central object of estimation in D-MRI analysis. We propose the Nearest-Neighbor Adaptive Regression Model (\texttt{NARM}), a spatially adaptive framework for FOD estimation that performs weighted local likelihood estimation over nested spatial neighborhoods, where the weights jointly encode spatial proximity and similarity among neighboring FODs, measured by either the optimal transport or Hellinger distance. To prevent over-smoothing while preserving structural heterogeneity, we introduce a voxel-wise rescaling scheme and a data-driven stopping rule based on minimum nearest-neighbor dissimilarity. We further develop a configuration-aware strategy for selecting the similarity-smoothing parameter, allowing the smoothing strength to adapt to local fiber complexity. Simulation studies demonstrate that \texttt{NARM} improves FOD estimation accuracy relative to voxel-wise methods and the existing spatial smoothing approach \texttt{PMARM}. Application to test--retest data from the Human Connectome Project additionally shows that \texttt{NARM} yields more reproducible FOD estimates. Implementation details and scripts for the simulation and real data analyses are available at \url{https://github.com/jie108/NARM}.
\end{abstract}

\noindent%
\vfill

\newpage
\spacingset{1.8} % DON'T change the spacing!

\section{Introduction}

Understanding brain anatomy and its relationship to function is a central problem in neuroscience. Diffusion-weighted magnetic resonance imaging (D-MRI) provides a noninvasive means of probing tissue microstructure by measuring water diffusion. In white matter, water diffuses preferentially along axonal bundles, allowing inference of local fiber orientations \citep{mori2007introduction}. Accurate estimation of fiber orientations at each voxel is crucial for tractography and structural connectivity analysis \citep{basser2000vivo, wong2016fiber}.

Early diffusion tensor imaging (DTI) methods model diffusion using a second-order tensor \citep{basser1994mr}, but they are unable to resolve complex intra-voxel configurations such as fiber crossings. High angular resolution diffusion imaging (HARDI) addresses this limitation by acquiring measurements along many gradient directions \citep{tuch2002high}. Among HARDI-based models, the fiber orientation distribution (FOD) \citep{tournier2004direct, tournier2007robust} directly characterizes both the number of fiber populations and their orientations within a voxel, making it particularly suitable for tractography.

FODs are symmetric spherical probability density functions and are often represented using spherical harmonics (SH). However, the global support of SH basis functions limits their ability to capture sharp, localized peaks. To overcome this limitation, \citet{yan2018estimating} proposed the \texttt{SN-lasso} method, which represents FODs using spherical needlets. Owing to their joint localization in the spatial and frequency domains \citep{ narcowich2006localized}, spherical needlets provide sparse and stable representations of FODs with sharp directional peaks.

Most existing FOD estimation methods, including \texttt{SN-lasso}, operate voxel-by-voxel and ignore spatial information. In practice, fiber orientations vary smoothly along tracts except at anatomical boundaries, suggesting that borrowing information from neighboring voxels can reduce variance. At the same time, regions with crossing fibers or tissue transitions exhibit abrupt changes, so naive spatial smoothing risks blurring distinct fiber populations. Effective spatial smoothing must therefore balance propagation in homogeneous regions with separation across heterogeneous boundaries.

The propagation–separation framework \citep{polzehl2000adaptive, polzehl2006propagation} provides a principled approach to this problem and has been adapted to D-MRI analysis in various forms, including spatially adaptive smoothing for DTI \citep{tabelow2008diffusion}, adaptive smoothing of HARDI images \citep{becker2012position, becker2014adaptive}, multiscale adaptive regression models \citep{li2011multiscale}, and the penalized multiscale adaptive regression model (\texttt{PMARM}) for ODF estimation \citep{rao2016sr}. Despite these advances, existing methods either rely on Euclidean similarity measures, use globally fixed smoothing parameters, or depend on stopping rules that are difficult to calibrate.

In this paper, we propose the \textit{Nearest-Neighbor Adaptive Regression Model} (\texttt{NARM}), a spatially adaptive framework for FOD estimation that integrates propagation–separation with voxel-wise \texttt{SN-lasso}. Rather than estimating each voxel independently, \texttt{NARM} performs weighted local likelihood estimation over expanding neighborhoods, where the weights adapt jointly to spatial proximity and similarity among neighboring FODs.

\texttt{NARM} introduces several methodological innovations. First, similarity is quantified using the Hellinger and optimal transport (OT) distances, which are more appropriate than Euclidean distance for comparing probability density functions. Second, we introduce a voxel-wise rescaling mechanism that mitigates both over-smoothing and under-smoothing during propagation. Third, \texttt{NARM} employs a fully data-driven, voxel-specific stopping rule based on minimum nearest-neighbor dissimilarity, enabling automatic adaptation to spatial heterogeneity. Finally, we adopt a configuration-aware strategy for selecting the similarity-smoothing parameter, allowing the degree of adaptivity to vary across different fiber types.

Simulation results show that \texttt{NARM} improves FOD estimation accuracy relative to voxel-wise \texttt{SN-lasso} and \texttt{PMARM}. We further analyze test--retest D-MRI data from 37 subjects in the WU--Minn Human Connectome Project (HCP) \citep{van2013wu}. Reproducibility is assessed by both the number of detected peaks, each corresponding to a major fiber bundle, and the optimal transport distance between FOD estimates at matched test--retest voxels. Across these measures, \texttt{NARM} consistently achieves higher reproducibility than competing approaches.

The remainder of the paper is organized as follows. Section~\ref{sec:voxel_FOD_estimation} introduces the FOD model and the voxel-wise \texttt{SN-lasso} estimator. Section~\ref{sec:spatial_FOD_estimation} presents \texttt{NARM}. Sections~\ref{sec:synthetic_experiments} and \ref{sec:real_data_experiments} report simulation and real-data results. Section~\ref{sec:conclusions} concludes with a brief discussion. Additional details are provided in the Supplementary Material.

\section{Voxel-wise FOD estimation}
\label{sec:voxel_FOD_estimation}

We briefly review the fiber orientation distribution (FOD) model \citep{tournier2004direct} and the voxel-wise \texttt{SN-lasso} estimator \citep{yan2018estimating}, which forms the basis of the \texttt{NARM} method.

%\subsection*{FOD model}

\begin{figure}[ht]
    \centering
    \includegraphics[width=0.8\textwidth]{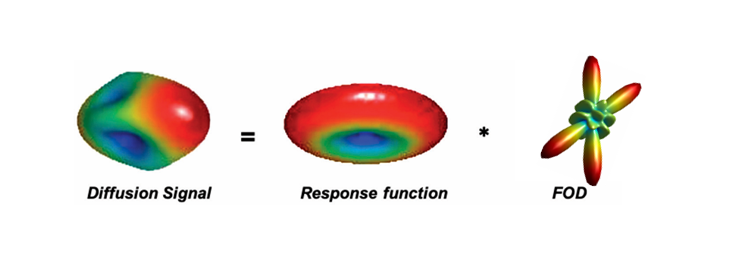}
    \caption{Illustration of the FOD model. The diffusion signal (left) is expressed as the spherical convolution of the response function (middle) and the FOD (right).}
    \label{fig:fod_model}
\end{figure}

At voxel $\bm{v}$, the diffusion signal $S(\bm{v},\cdot)$ is modeled as the spherical convolution of the FOD $F(\bm{v},\cdot)$ and a response function $R(\cdot)$:
\begin{equation}
\label{eq:spherical_convolution_model}
S(\bm{v}, \bm{x}) 
=
\int_{\mathbb{S}^2}
R(\bm{x}^\top \bm{y})\,
F(\bm{v}, \bm{y})
\, d\omega(\bm{y}),
\qquad \bm{x}\in\mathbb{S}^2 .
\end{equation}
Here $F(\bm{v},\cdot)$ is a symmetric spherical probability density function describing fiber orientations, and $R(\cdot)$ is an axially symmetric response kernel associated with a single coherently aligned fiber population. The response function is assumed to be invariant across voxels and is pre-estimated; throughout, it is treated as known (Figure~\ref{fig:fod_model}). 
The diffusion-weighted measurements $\bm{y}(\bm{v})=\{y_i(\bm{v})\}_{i=1}^n$ are observed along gradient directions $\{\bm{x}_i\}_{i=1}^n \subset \mathbb{S}^2$, where $y_i(\bm{v})$ is a noisy observation of the diffusion signal $S(\bm{v},\bm{x}_i)$.
The goal is to estimate $F(\bm{v},\cdot)$ from $\bm{y}(\bm{v})$.

Let $\{\Phi_{lm}\}$ denote the real symmetric spherical harmonic (SH) basis \citep{descoteaux2007regularized}. Restricting to degrees $0 \le l \le l_{\max}$ yields
$L=\frac{(l_{\max}+1)(l_{\max}+2)}{2}$ basis functions for representing $S(\bm{v},\cdot)$, $R(\cdot)$, and $F(\bm{v},\cdot)$.
Under model \eqref{eq:spherical_convolution_model}, the measurements satisfy
\begin{equation}
\label{eq:SH_regression}
\bm{y}(\bm{v})
=
\bm{\Phi}\bm{R}\bm{f}(\bm{v})
+
\bm{\epsilon},
\end{equation}
where $\bm{\Phi}$ is the $n\times L$ SH design matrix evaluated at the gradient directions $\{\bm{x}_i\}_{i=1}^n$, $\bm{R}$ is an $L\times L$ block-diagonal matrix with diagonal entries $\sqrt{4\pi/(2l+1)}\,r_l$, and $\bm{r}=(r_l)$ and $\bm{f}(\bm{v})$ are the SH coefficient vectors of the response function $R(\cdot)$ and the FOD $F(\bm{v},\cdot)$, respectively.
%\subsection*{Spherical needlet representation and \texttt{SN-lasso}}

While convenient, SH bases have global support and are inefficient for representing FODs with sharp, localized peaks. To better capture such localized structure, \citet{yan2018estimating} proposed the \texttt{SN-lasso} estimator, which represents FODs using symmetric spherical needlets (SN) that are jointly localized in space and frequency \citep{narcowich2006localized}. 

Let $j_{\max}=\lceil\log_2(l_{\max})\rceil$ and 
$N=2^{2j_{\max}+3}-1$ denote the number of SN basis functions up to level $j_{\max}$. The SH coefficients admit the representation
\[
\bm{f}(\bm{v})=\bm{C}\bm{\beta}(\bm{v}),
\]
where $\bm{\beta}(\bm{v})$ are the SN coefficients and $\bm{C}$ is the transformation matrix from SN to SH bases. Substituting this representation into \eqref{eq:SH_regression} yields
\begin{equation}
\label{eq:SN_regression}
\bm{y}(\bm{v})
=
\bm{\Phi}\bm{R}\bm{C}\bm{\beta}(\bm{v})
+
\bm{\epsilon}.
\end{equation}

To exploit the sparsity of the SN coefficients, \texttt{SN-lasso} estimates $\bm{\beta}(\bm{v})$ by solving
\begin{equation}
\label{eq:SN_lasso}
\hat{\bm{\beta}}(\bm{v})
=
\arg\min_{\bm{\beta}:\,\tilde{\bm{\Phi}}\bm{C}\bm{\beta}\succeq 0}
\frac{1}{2}
\left\|
\bm{y}(\bm{v})
-
\bm{\Phi}\bm{R}\bm{C}\bm{\beta}
\right\|_2^2
+
\lambda\|\bm{\beta}\|_1,
\end{equation}
where $\lambda \ge 0$ controls the degree of sparsity, $\tilde{\bm{\Phi}}$ is the SH evaluation matrix on a dense spherical grid, and the constraint
$\tilde{\bm{\Phi}}\bm{C}\bm{\beta}\succeq 0$ enforces nonnegativity of the estimated FOD on that grid.

The optimization problem is solved using ADMM \citep{boyd2011distributed}. The tuning parameter $\lambda$ is selected as the value at which the residual sum of squares stabilizes. Peak detection is subsequently performed to identify dominant fiber orientations.
%%%%%%%%%%%%%%%%
%%%%%%%%%%%%%%%%
\section{Spatially-smoothed FOD estimation}
\label{sec:spatial_FOD_estimation}

\subsection{Spatially adaptive smoothing framework}
\label{subsec:spatial_framework}

\begin{figure}[ht]
    \centering
    \includegraphics[width=3.5in]{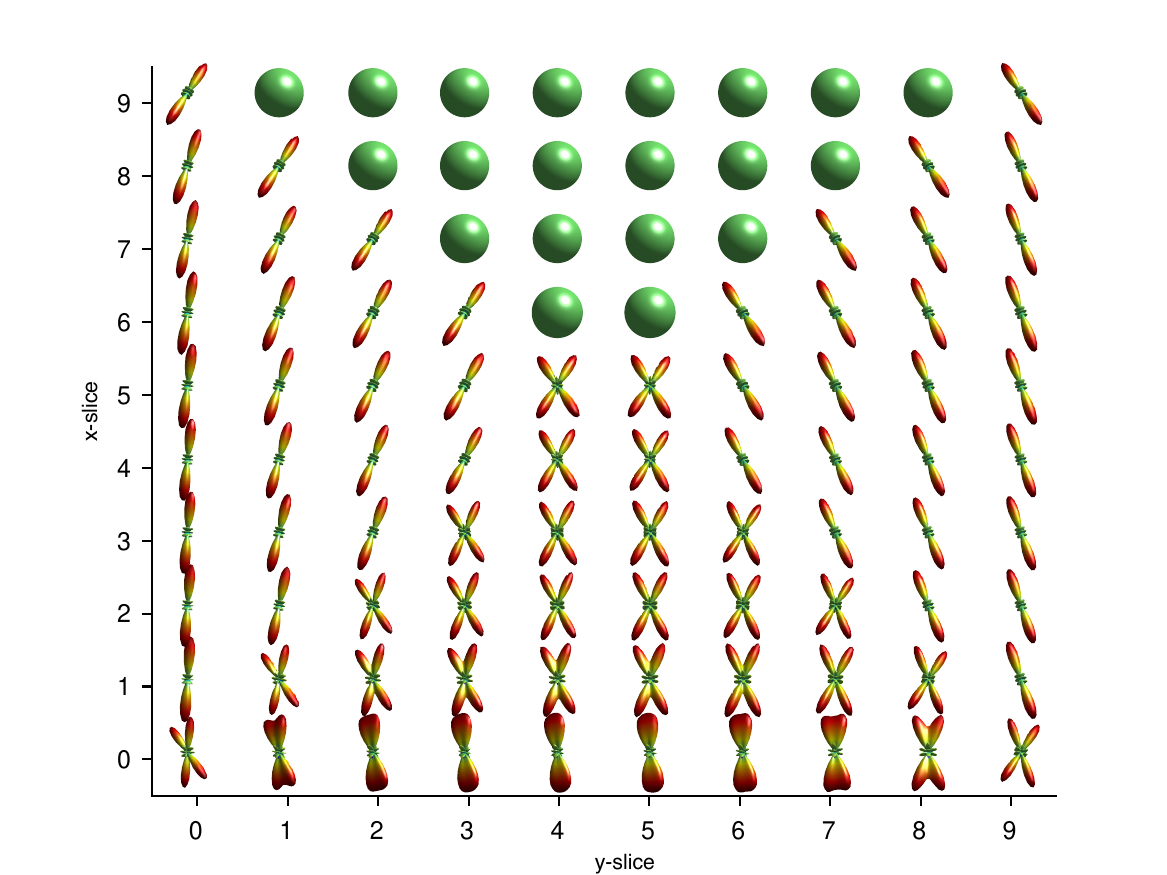}
  \caption{Simulated FODs at $z$-slice $=5$ from a $10 \times 10 \times 10$ three-dimensional grid. Green spheres denote isotropic voxels (uniform FODs with no preferred fiber orientation). For example, the voxel at $(x,y)=(9,0)$ exhibits a single-fiber FOD, whereas the voxel at $(x,y)=(5,4)$ exhibits a two-fiber FOD.}
    \label{fig:simu-fod}
\end{figure}

The voxel-wise \texttt{SN-lasso} estimator uses only $\bm{y}(\bm{v})$ to estimate the FOD at voxel $\bm{v}$. In white matter, however, fiber bundles vary smoothly across voxels along their principal directions, suggesting that borrowing information from nearby voxels can reduce estimation variance. At the same time, spatial homogeneity varies substantially: it may extend over large regions in coherent tracts but breaks down locally near crossings or tissue boundaries, as illustrated in Figure~\ref{fig:simu-fod}.

To balance propagation in homogeneous regions against separation across discontinuities, we propose the \textit{Nearest-Neighbor Adaptive Regression Model} (\texttt{NARM}). The method combines local likelihood ideas \citep{fan1996local} with the propagation--separation principle \citep{polzehl2000adaptive, polzehl2006propagation}: it iteratively enlarges the spatial neighborhood (propagation) while adaptively downweighting dissimilar neighbors and terminating early when local homogeneity no longer holds (separation; see Sections~\ref{subsec:weights} and~\ref{subsection:stop}).

Throughout, $\bm{v}$ denotes both a voxel and its grid index on the ROI lattice, with the origin located at the lower-left corner (indexed as $(0,0,0)$). Let $d_s$ denote the neighborhood radius at step $s$. Define
$0=d_0<d_1<\cdots<d_S$ with $d_s=r^s$ for $s=1,\ldots,S$, where $r>1$ is the neighborhood expansion rate and $S$ is the maximum number of steps.

On a Cartesian grid with unit spacing, define the nested spherical neighborhoods
\[
\mathcal{N}^{(s)}(\bm{v})
:=
\{\bm{v}' : \|\bm{v}-\bm{v}'\|_2 \le d_s\},
\qquad s=0,1,\ldots,S.
\]

At step $s$, let $\omega^{(s)}(\bm{v},\bm{v}')\ge 0$ be weights satisfying
\[
\sum_{\bm{v}'\in \mathcal{N}^{(s)}(\bm{v})}\omega^{(s)}(\bm{v},\bm{v}')=1.
\]
We estimate the spherical needlet coefficients at $\bm{v}$ by the weighted regression
\begin{equation}
\label{eq:weighted_SN_lasso}
\hat{\bm{\beta}}^{(s)}(\bm{v})
:=
\arg\min_{\bm{\beta}:\,\tilde{\bm{\Phi}}\bm{C}\bm{\beta}\succeq \bm{0}}
\sum_{\bm{v}'\in \mathcal{N}^{(s)}(\bm{v})}
\omega^{(s)}(\bm{v},\bm{v}')\,
\frac{1}{2}\left\|\bm{y}(\bm{v}')-\bm{\Phi}\bm{R}\bm{C}\bm{\beta}\right\|_2^2
+\lambda\|\bm{\beta}\|_1 ,
\end{equation}
or equivalently,
\[
\hat{\bm{\beta}}^{(s)}(\bm{v})
=
\arg\min_{\bm{\beta}:\,\tilde{\bm{\Phi}}\bm{C}\bm{\beta}\succeq \bm{0}}
\frac{1}{2}\left\|\bm{y}^{(s)}(\bm{v})-\bm{\Phi}\bm{R}\bm{C}\bm{\beta}\right\|_2^2
+\lambda\|\bm{\beta}\|_1 ,
\quad
\bm{y}^{(s)}(\bm{v})
:=
\sum_{\bm{v}'\in \mathcal{N}^{(s)}(\bm{v})}\omega^{(s)}(\bm{v},\bm{v}')\,\bm{y}(\bm{v}').
\]
The step-$s$ FOD estimate (evaluated on the grid) is
\[
\hat{\bm{F}}^{(s)}(\bm{v})
:=
\tilde{\bm{\Phi}}\,\bm{C}\,\hat{\bm{\beta}}^{(s)}(\bm{v}).
\]
At $s=0$, $\mathcal{N}^{(0)}(\bm{v})=\{\bm{v}\}$, and \eqref{eq:weighted_SN_lasso} reduces to the voxel-wise \texttt{SN-lasso} estimator \eqref{eq:SN_lasso}.

We set $r=1.15$, $S=10$ for the simulation experiments, and $S=20$ for the HCP application. We further use $l_{\max}=8$ and $j_{\max}=3$, corresponding to $L=45$ SH basis functions and $N=511$ SN basis functions, with $\tilde{\bm{\Phi}}$ constructed using $642$ grid points from a level-4 icosphere mesh.

\subsection{Adaptive weights}
\label{subsec:weights}

The weights $\omega^{(s)}(\bm{v},\bm{v}')$ implement propagation--separation via two components: a spatial kernel that favors nearby voxels, and a similarity kernel that downweights structurally dissimilar neighbors based on the previous-step FOD estimates, thereby preventing smoothing across fiber crossings or tissue boundaries. As illustrated in Figure~\ref{fig:simu-fod}, a voxel may have both similar and markedly different neighbors. For example, this occurs for the voxel at $(x,y)=(5,4)$.

For $s\ge 1$, define
\begin{equation}
\label{eq:weight}
\omega^{(s)}(\bm{v}, \bm{v}')
:=
\frac{
K_{\mathrm{loc}}\!\left(\|\bm{v}-\bm{v}'\|_2/d_s\right)\,
K_{\mathrm{sim}}\!\left(\gamma \cdot \mathrm{Dist}_{s-1}(\bm{v},\bm{v}')\right)
}{
\sum_{\bm{\tilde v}\in \mathcal{N}^{(s)}(\bm{v})}
K_{\mathrm{loc}}\!\left(\|\bm{v}-\bm{\tilde v}\|_2/d_s\right)\,
K_{\mathrm{sim}}\!\left(\gamma\cdot\mathrm{Dist}_{s-1}(\bm{v},\bm{\tilde v})\right)
},
\end{equation}
where $\gamma$ controls the strength of similarity adaptation, and
\[
\mathrm{Dist}_{s-1}(\bm{v},\bm{v}')
:=
\mathrm{Dist}\!\left(
\hat{\bm{F}}^{(s-1)}(\bm{v}),
\hat{\bm{F}}^{(s-1)}(\bm{v}')
\right).
\]

We next specify $K_{\mathrm{loc}}(\cdot)$, $K_{\mathrm{sim}}(\cdot)$, $\mathrm{Dist}(\cdot, \cdot)$, and a rescaling scheme.

\subsubsection*{Spatial kernel}

We use the compactly supported kernel
\[
K_{\mathrm{loc}}(u)=(1-u^2)_+ .
\]
Thus, $K_{\mathrm{loc}}(\|\bm{v}-\bm{v}'\|_2/d_s)$ assigns larger weights to nearby voxels and is supported on $\mathcal{N}^{(s)}(\bm{v})$. As $d_s$ increases, the support expands and the decay with physical distance becomes flatter, such that distant voxels incur less spatial penalty and the weights become more uniform, enabling broader propagation.

\subsubsection*{Similarity kernel}
We use
\[
K_{\mathrm{sim}}(u)=\exp(-u^2).
\]
Accordingly, $K_{\mathrm{sim}}(\gamma \cdot \mathrm{Dist}_{s-1}(\bm{v},\bm{v}'))$ downweights neighbors whose estimated FODs differ substantially from that of $\bm{v}$, thereby inducing separation across heterogeneous regions. The smoothing parameter $\gamma\ge 0$ controls the strength of this effect: $\gamma=0$ yields pure spatial smoothing, whereas larger values of $\gamma$ induce stronger separation across dissimilar neighbors, with $\gamma\to\infty$ recovering voxel-wise estimation.

\subsubsection*{Dissimilarity measure}

Let $\bm{p}^{(s)}(\bm{v})$ denote the unit-sum normalization of $\hat{\bm{F}}^{(s)}(\bm{v})$. We consider two dissimilarity measures. The first is the \textit{Hellinger distance}
\[
\mathrm{Dist}_s(\bm{v},\bm{v}')
=
\frac{1}{\sqrt{2}}
\left\|
\sqrt{\bm{p}^{(s)}(\bm{v})}-\sqrt{\bm{p}^{(s)}(\bm{v}')}
\right\|_2.
\]

The second is an entropy-regularized \textit{optimal transport (OT) distance} computed via the Sinkhorn algorithm \citep{Villani2009, cuturi2013sinkhorn}:
\[
\mathrm{Dist}_s(\bm{v},\bm{v}')
=
d_{OT}^{\kappa}\!\left(\bm{p}^{(s)}(\bm{v}), \bm{p}^{(s)}(\bm{v}')\right),
\]
where $\kappa \ge 0$ is a regularization parameter. Smaller values of $\kappa$ yield a closer approximation to the unregularized OT distance, at the expense of increased computational cost, while $\kappa=0$ corresponds to the unregularized OT distance. In the numerical studies, we use $\kappa=0.02$ for OT computation within the \texttt{NARM} estimation procedure and $\kappa=0.005$ for OT-based performance evaluation. A detailed definition of $d_{OT}^{\kappa}$ is provided in Supplementary Material Section~\ref{sec:OT}.

The Hellinger distance is computationally efficient, whereas the OT distance better reflects angular displacement and is often more faithful for comparing FODs with shifted or partially misaligned peaks.

\subsubsection*{Rescaling scheme}
\label{subsec:rescale}

The weights in \eqref{eq:weight} may become suboptimal when a voxel is already well smoothed, in which case further smoothing increases bias, or when it remains severely under-smoothed, in which case stronger smoothing is needed. We therefore rescale the smoothing parameter $\gamma$ using a local homogeneity proxy based on the estimates from the previous step.

For $s\ge 0$, define the minimum nearest-neighbor dissimilarity
\begin{equation}
\label{eq:MNN}
\text{MNN-Dist}_s(\bm{v})
:=
\min_{\bm{v}'\in\mathcal{A}(\bm{v})}
\mathrm{Dist}_{s}(\bm{v},\bm{v}'),
\end{equation}
where $\mathcal{A}(\bm{v})$ denotes the set of six face-adjacent neighbors of $\bm{v}$. For a small $\alpha>0$, let $\text{MNN-Dist}_s^{\alpha}$ and $\text{MNN-Dist}_s^{1-\alpha}$ denote the $100\alpha$-th and $100(1-\alpha)$-th percentiles of $\text{MNN-Dist}_s(\bm{v})$ across voxels. Voxels with $\text{MNN-Dist}_s(\bm{v}) > \text{MNN-Dist}_s^{1-\alpha}$ are regarded as under-smoothed, whereas those with $\text{MNN-Dist}_s(\bm{v}) < \text{MNN-Dist}_s^{\alpha}$ are regarded as well smoothed. In our implementation, we set $\alpha=0.15$.

For $s \ge 1$, define
\[
\gamma^{(s)}(\bm{v})
:=
\min\!\left\{\frac{\text{MNN-Dist}_{s-1}^{1-\alpha}}{\text{MNN-Dist}_{s-1}(\bm{v})},\,1\right\}
\cdot
\max\!\left\{\frac{\text{MNN-Dist}_{s-1}^{\alpha}}{\text{MNN-Dist}_{s-1}(\bm{v})},\,1\right\}.
\]
We then compute the weights as in \eqref{eq:weight}, replacing the global smoothing parameter $\gamma$ by the voxel-specific quantity $\gamma\cdot\gamma^{(s)}(\bm{v})$. Thus, for under-smoothed voxels, $\gamma^{(s)}(\bm{v})<1$ yields more aggressive smoothing, whereas for well-smoothed voxels, $\gamma^{(s)}(\bm{v})>1$ reduces the degree of smoothing. For the remaining approximately $(1-2\alpha)\cdot 100\%$ of voxels, $\gamma^{(s)}(\bm{v})=1$, and thus no rescaling is applied.

\subsection{Stopping rule}
\label{subsection:stop}

To prevent over-smoothing, updates at a voxel should stop once enlarging the neighborhood no longer improves local homogeneity. We use two complementary criteria.

First, if the peak detection procedure identifies voxel $\bm{v}$ as isotropic (i.e., having zero peaks) at any iteration, updates at that voxel are terminated, as false positive identification of isotropic voxels is empirically rare and such voxels require no further smoothing. %we stop updating that voxel, since false positive identification of isotropic voxels is empirically rare.

Second, using $\text{MNN-Dist}_s(\bm{v})$ from \eqref{eq:MNN}, we terminate updates at $\bm{v}$ when local homogeneity ceases to improve. Specifically, for $s \ge 2$, if
\[
\min\!\left\{\text{MNN-Dist}_s(\bm{v}),\,\text{MNN-Dist}_{s-1}(\bm{v})\right\}
\ge
\text{MNN-Dist}_{s-2}(\bm{v}),
\]
the estimate is frozen at
%we stop and set
\[
\hat{\bm{F}}^{(s')}(\bm{v}) \equiv \hat{\bm{F}}^{(s-2)}(\bm{v}),
\qquad \forall\, s'\ge s-1.
\]
Thus, different voxels may stop at different steps, naturally adapting to spatial heterogeneity.

\subsection{Data-driven choice of $\gamma$ and two-stage procedure}

Voxels are classified into three categories according to their fiber configurations: (i) isotropic, corresponding to a uniform FOD; (ii) single-fiber; and (iii) multi-fiber. See Figure~\ref{fig:simu-fod} for examples of each category. Preliminary simulation results indicate a trade-off between adaptivity and robustness. With the OT distance, \texttt{NARM} substantially improves identification of isotropic voxels, which would otherwise often be misidentified as having multiple spurious peaks. This improvement is driven by similarity weighting: FODs with many spurious peaks tend to be closer, in OT distance, to the uniform distribution than to FODs with a few well-separated peaks, thereby enabling effective information sharing across neighboring isotropic voxels.

In contrast, multi-fiber FOD estimation is inherently more difficult, especially when diffusion contrast is low (e.g., at small $b$-values) or the noise level is high. Note that the $b$-value, measured in s/mm$^2$, controls the strength of diffusion weighting with smaller $b$ yielding lower angular contrast. In these settings, increased estimation variability degrades the dissimilarity measure and hence the adaptive weights. Consequently, pure spatial smoothing ($\gamma=0$) is often more stable. However, applying pure spatial smoothing from the outset can hinder isotropic identification when some isotropic voxels are initially mislabeled as multi-fiber. Single-fiber voxels lie between these two extremes and typically require only mild smoothing.

We therefore adopt a data-driven choice of the smoothing parameter $\gamma$ based on nearest-neighbor dissimilarities, which capture local geometry using only pairwise distances \citep{chen2017simple}. This approach has the additional advantage that $\gamma$ is automatically adapted to the scale of the underlying distance metric. We also introduce an optional two-stage refinement procedure.

\medskip
\noindent\textbf{Stage 1: Adaptive smoothing for voxel classification.}
Let $np(\bm{v})$ denote the number of detected fibers from the voxel-wise \texttt{SN-lasso} estimate $\hat{\bm{F}}^{(0)}(\bm{v})$, and define
\[
\mathcal{C}_0 := \{\bm{v} : np(\bm{v}) = 0\}, \quad
\mathcal{C}_1 := \{\bm{v} : np(\bm{v}) = 1\}, \quad
\mathcal{C}_2 := \{\bm{v} : np(\bm{v}) \ge 2\}.
\]
Set
\[
\gamma_1 :=
\frac{1}{\operatorname{median}_{\bm{v}\in \mathcal{C}_1}\text{MNN-Dist}_0(\bm{v})},
\qquad
\gamma_2 :=
\frac{1}{\operatorname{median}_{\bm{v}\in \mathcal{C}_2}\text{MNN-Dist}_0(\bm{v})},
\]
where $\text{MNN-Dist}_0(\bm{v})$ is computed from the voxel-wise \texttt{SN-lasso} estimates as defined in \eqref{eq:MNN}. We set $\gamma=\gamma_1$ for voxels in $\mathcal{C}_1$ and $\gamma=\gamma_2$ for voxels in $\mathcal{C}_2$. In practice, we observe that $\gamma_1$ is typically about $2.5\times \gamma_2$.

Under our stopping criterion, voxels in $\mathcal{C}_0$ are not updated and therefore remain isotropic; that is, their FODs stay unchanged throughout the procedure.

\medskip
\noindent\textbf{Stage 2: Configuration-specific refinement (optional).}
When diffusion contrast is low (e.g., $b \approx 1000\,\mathrm{s/mm}^2$) or the noise level is high, we can further refine the estimates in a second stage. Specifically, after Stage~1, we update the voxel categories as
\[
\widetilde{\mathcal{C}}_0 := \{\bm{v} : \bm{v}\text{ is identified as isotropic by the Stage 1 estimate}\}, \quad
\widetilde{\mathcal{C}}_1 := \mathcal{C}_1 \setminus \widetilde{\mathcal{C}}_0, \quad
\widetilde{\mathcal{C}}_2 := \mathcal{C}_2 \setminus \widetilde{\mathcal{C}}_0.
\]
We then recalibrate the single-fiber smoothing parameter as
\[
\widetilde{\gamma}_1
:=
\frac{1}{\operatorname{median}_{\bm{v}\in \widetilde{\mathcal{C}}_1}\text{MNN-Dist}_0(\bm{v})}.
\]
In Stage~2, we set $\gamma=\widetilde{\gamma}_1$ for voxels in $\widetilde{\mathcal{C}}_1$ and $\gamma=0$ (i.e., pure spatial smoothing) for voxels in $\widetilde{\mathcal{C}}_2$; voxels in $\widetilde{\mathcal{C}}_0$ are not updated. The stopping rule in Section~\ref{subsection:stop} is applied throughout.

We refer to Stage~1 as \texttt{NARM-OT-1}. If Stage~2 uses the OT distance, the full method is \texttt{NARM-OT}; if it uses the Hellinger distance, it is \texttt{NARM-H}.

\subsection{Comparison with \texttt{PMARM}}

\citet{rao2016sr} proposed the \textit{Penalized Multiscale Adaptive Regression Model} (\texttt{PMARM}) for spatially adaptive estimation of the orientation distribution function (ODF). Since both ODFs and FODs are spherical densities, \texttt{PMARM} can be readily adapted to FOD estimation. Like \texttt{NARM}, it forms weighted averages over expanding neighborhoods and uses a stopping rule to control over-smoothing. The two methods differ in several key respects.

First, \texttt{PMARM} uses an $\ell_2$ (Euclidean) distance, whereas \texttt{NARM} uses the Hellinger or OT distance, both of which are tailored to comparing probability distributions and better reflect the structure of FODs. Second, \texttt{NARM} includes voxel-wise rescaling to mitigate both over-smoothing and under-smoothing during propagation. Third, \texttt{PMARM} adopts a global stopping rule: for a pre-specified constant $c_0>0$, smoothing at voxel $\bm{v}$ is terminated if
\[
\|\hat{\bm{F}}^{(s)}(\bm{v})-\hat{\bm{F}}^{(s-1)}(\bm{v})\|_2
>
c_s:=c_0 \cdot \chi^2_{(1)}(0.6/s),
\]
where $\chi^2_{(1)}(\alpha)$ denotes the $100(1-\alpha)$-th percentile of the $\chi^2_{(1)}$ distribution. The rationale is that, if the distance between the estimates at steps $s$ and $s-1$ exceeds $c_s$, then some ``bad'' signals from neighboring voxels may have been incorporated, leading to a substantial change in the estimated FOD; in that case, the smoothing procedure is terminated.

However, the use of a chi-square quantile has no clear justification in this context, and the performance can be sensitive to the choice of $c_0$. In contrast, \texttt{NARM} uses a voxel-wise, data-driven stopping rule based on nearest-neighbor dissimilarity, which better adapts to heterogeneous regions, such as coherent tracts and fiber crossings. Finally, \texttt{PMARM} fixes the similarity smoothing parameter $\gamma$ a priori, whereas \texttt{NARM} selects $\gamma$ in a configuration-aware, data-driven manner.

In the numerical studies, we use the same neighborhood expansion settings for \texttt{PMARM} as for \texttt{NARM}, namely $r=1.15$, with $S=10$ for the simulation experiments and $S=20$ for the HCP application. We evaluate several combinations of tuning parameters $(c_0,\gamma)$ and report results for two representative choices that yield overall competitive performance relative to \texttt{NARM}.

\section{Simulation experiments}\label{sec:synthetic_experiments}

In this section, we use synthetic D-MRI data to evaluate the performance of the proposed \texttt{NARM} method and compare it with the voxel-wise \texttt{SN-lasso} and \texttt{PMARM} methods.

We generate true fiber directions and the corresponding diffusion-weighted measurements on a $10 \times 10 \times 10$ three-dimensional grid, with fiber directions varying smoothly across voxels. Each voxel contains either one dominant fiber direction ($419$ voxels), two crossing fiber directions ($343$ voxels), or no fiber direction ($238$ voxels), corresponding to isotropic diffusion.

For voxels containing $K\ge 1$ fiber bundles, the true FOD is defined as a weighted sum of Dirac masses:
\[
F(\theta,\phi)=\sum_{k=1}^K p_k \,\delta_{(\theta_k,\phi_k)}(\theta,\phi),
\qquad \theta\in[0,\pi],\ \phi\in[0,2\pi),
\]
where $p_k$ is the volume fraction of the $k$th fiber bundle and $(\theta_k,\phi_k)$ is the spherical coordinate of its direction. For voxels with two fiber bundles, we set $p_1=p_2=0.5$. For isotropic voxels, the true FOD is uniform on the sphere:
\[
F(\theta,\phi)=\frac{1}{4\pi}.
\]
An illustration of the true FODs at $z$-slice = 5 is provided in Figure~\ref{fig:simu-fod} .

The response function is generated from the single-tensor model \citep{le1995diffusion, basser2002diffusion, mori2007introduction}, as detailed in Supplementary Material Section~\ref{subsec:supp_single_tensor_model}:
\[
R(\cos\theta)
:=
S_0 \exp\!\left\{-b\left(\bar{\lambda}\cos^2\theta+\underline{\lambda}\sin^2\theta\right)\right\},
\qquad \theta\in[0,\pi].
\]

where $S_0=1$, the $b$-value is set to either $1000\,\mathrm{s/mm}^2$ or $3000\,\mathrm{s/mm}^2$, and $\underline{\lambda}=10^{-3}\,\mathrm{mm}^2/\mathrm{s}$. 
For voxels containing at least one fiber bundle, we set $\bar{\lambda}/\underline{\lambda}=10$, whereas for isotropic voxels we set $\bar{\lambda}/\underline{\lambda}=1$.

Given the true FOD and response function, we generate noiseless diffusion-weighted signals along $n$ gradient directions according to the FOD model \eqref{eq:spherical_convolution_model}. The gradient directions are taken from icosphere meshes on the half-sphere, with $n=41$ or $81$. Independent Rician noise \citep{gudbjartsson1995rician, hahn2006random} is then added to obtain the observed diffusion-weighted measurements. The signal-to-noise ratio (SNR), defined as $S_0/\sigma$, with $\sigma$ denoting the Rician noise level, is set to $10$ or $20$. These settings reflect typical D-MRI acquisition protocols, including those used in the Alzheimer's Disease Neuroimaging Initiative (ADNI) and the Human Connectome Project (HCP).

\begin{figure}[ht]
    \centering
    \includegraphics[width=1\textwidth]{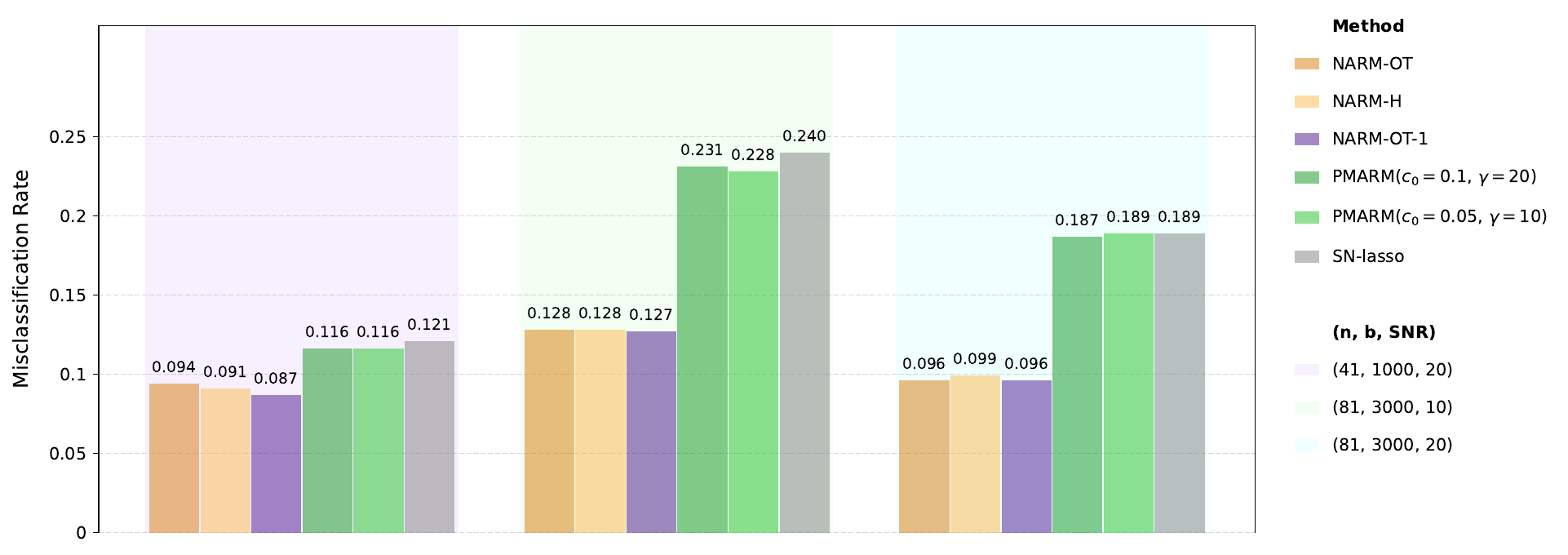}
\caption{Simulation results: overall misclassification rates across all voxels for different methods and simulation settings $(n,b,\mathrm{SNR})$.}
\label{fig:simu-mis}
\end{figure}

Performance is evaluated using the following metrics:

(i) The overall misclassification rate across all voxels, defined as the proportion of voxels assigned to an incorrect category (e.g., an isotropic voxel may be misidentified as a single-fiber voxel), is reported in Figure~\ref{fig:simu-mis}. More detailed results, including over- and under-identification rates within each voxel category, are provided in Supplementary Tables~\ref{tab:simu-tab1}--\ref{tab:simu-tab3}.

\begin{figure}[ht]
    \centering
    \includegraphics[width=1\textwidth]{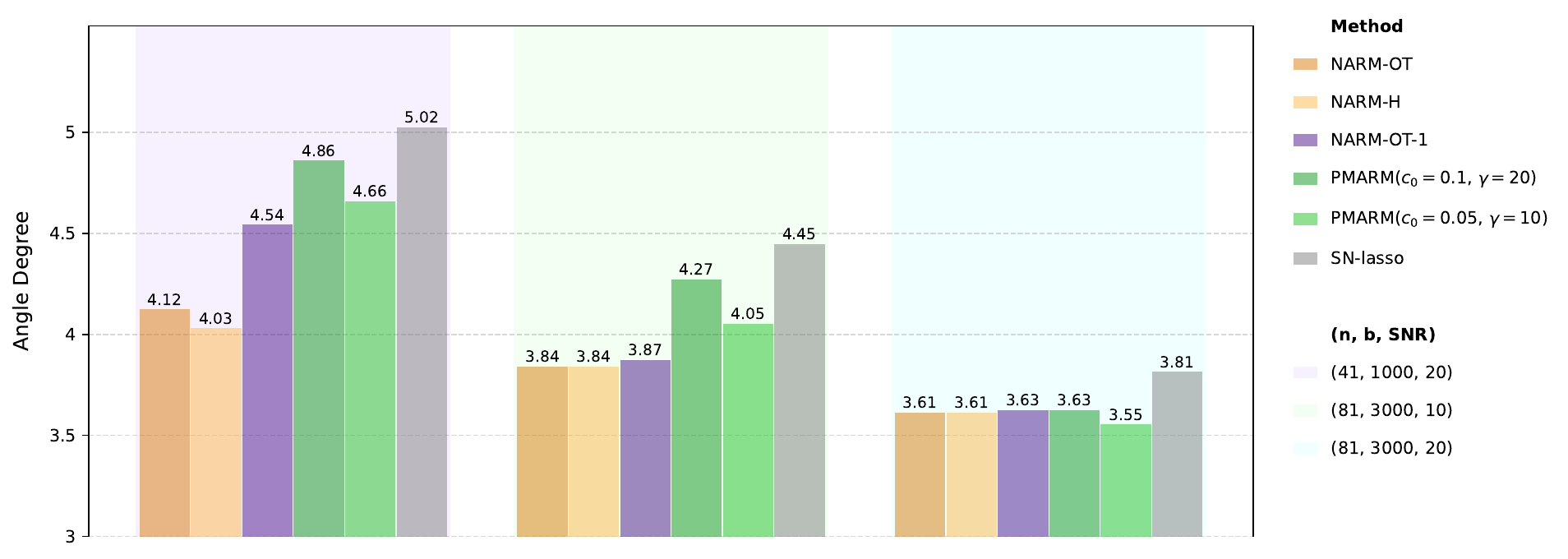}
\caption{Simulation results: median angular error on anisotropic voxels correctly identified by all competing methods under each simulation setting $(n,b,\mathrm{SNR})$.}
\label{fig:simu-angle}
\end{figure}

(ii) The angular error, evaluated on anisotropic voxels (i.e., single-fiber or two-fiber voxels) correctly identified by all competing methods (between $639$ and $683$ such voxels across the three $(n,b,\mathrm{SNR})$ settings), is reported in Figure~\ref{fig:simu-angle}.

As shown in these figures and tables, the one-stage \texttt{NARM-OT-1} achieves the lowest misclassification rates, closely followed by the two-stage \texttt{NARM-OT} and \texttt{NARM-H} variants, whereas \texttt{PMARM} and voxel-wise \texttt{SN-lasso} exhibit substantially higher misclassification rates.

For angular error, all smoothing-based methods perform comparably and outperform voxel-wise \texttt{SN-lasso} in the strong-signal, high-angular-contrast setting $(n,b,\mathrm{SNR})=(81,3000,20)$. In the lower-signal or noisier settings, $(41,1000,20)$ and $(81,3000,10)$, the two-stage \texttt{NARM-OT} and \texttt{NARM-H} achieve the smallest angular errors, followed by the one-stage \texttt{NARM-OT-1}. Furthermore, the improvement of the two-stage \texttt{NARM} methods over the one-stage method is most pronounced in the low-$b$-value setting $(n,b,\mathrm{SNR})=(41,1000,20)$.

\begin{figure}[ht]
    \centering
    \includegraphics[width=1\textwidth]{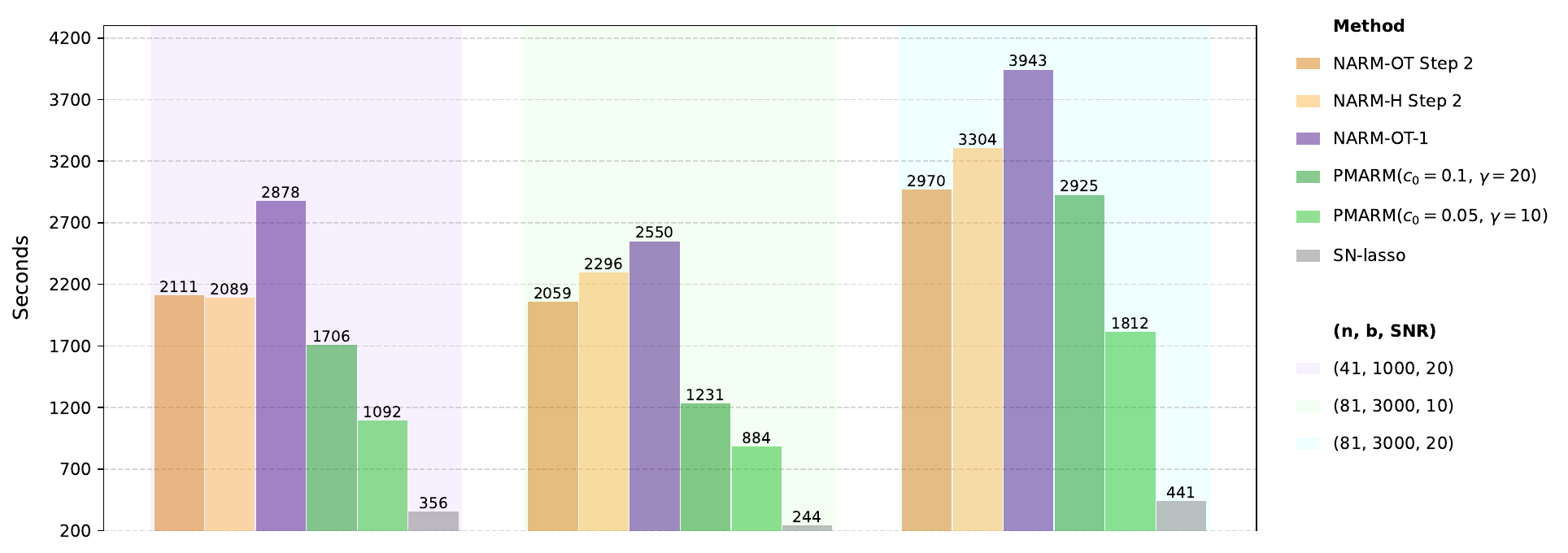}
\caption{Simulation results: total running time (seconds) for each method under different simulation settings $(n,b,\mathrm{SNR})$. The running time for \texttt{SN-lasso} reflects voxel-wise estimation only.}
\label{fig:simu-runtime}
\end{figure}

In addition, Figure~\ref{fig:simu-runtime} reports the total running time. Overall, each stage of \texttt{NARM} is typically about $1.5$ to $3$ times as computationally expensive as \texttt{PMARM}, although the ratio can be close to $1$ under some settings. This difference is partly attributable to the stopping rule. For example, under the setting $(n,b,\mathrm{SNR})=(81,3000,20)$, after step $s=5$, about $45\%$ of voxels have stopped under \texttt{NARM-OT-1} and about $60\%$ under the second stage of \texttt{NARM-OT} and \texttt{NARM-H}, whereas the corresponding proportions are $78\%$ and $96\%$ for \texttt{PMARM} with $(c_0=0.1,\gamma=20)$ and $(c_0=0.05,\gamma=10)$, respectively.
%At the final step $s=S=10$, about $75\%$ to $85\%$ of voxels have stopped under the \texttt{NARM} methods, compared with more than $97\%$ under the \texttt{PMARM} methods.

\section{HCP application}\label{sec:real_data_experiments}

In this section, we evaluate \texttt{NARM}, \texttt{PMARM}, and \texttt{SN-lasso} using test--retest data from the WU--Minn Human Connectome Project (HCP) \citep{van2013wu}. Specifically, we use D-MRI measurements acquired at $b=3000\,\mathrm{s/mm}^2$ with $91$ gradient directions and voxel size $1.25\times 1.25\times 1.25\,\mathrm{mm}^3$, from 37 healthy young adults aged 22--35. Details of data preprocessing and response function estimation are provided in Supplementary Material Sections~\ref{subsec:hcp_preprocessing} and \ref{subsec:hcp_response}.

We focus on a region containing crossing fibers of the \textit{Corticospinal Tract} (CST) and the \textit{Superior Longitudinal Fasciculus} (SLF) in MNI152-T1 template space. Using \textit{FSLeyes} \citep{fsleyes} and the \textit{JHU White-Matter Tractography Atlas} \citep{WAKANA2007,HUA2008}, we construct ROI masks in the left hemisphere. The ROI consists of voxels with $X\in[-40,-22]$, $Y\in[-30,-12]$, and $Z\in[22,40]$ in MNI152 space as shown in Figure~\ref{fig:registration} (top panel).

Using the inverse transformation from the registration step, we map these masks back to each subject's native space as shown in Figure~\ref{fig:registration} (bottom panel). In native space, the numbers of voxels within the ROI in the test and retest scans have means (standard deviations) of 3906.1 (493.0) and 3912.3 (477.5), respectively. The paired difference in voxel counts has a 95\% confidence interval of $(-32.1,\,19.8)$.

We apply \texttt{SN-lasso}, \texttt{NARM}, and \texttt{PMARM} to estimate the FODs for  voxels within the ROI in each subject's native space. Since ground truth is unavailable, performance is assessed via test--retest reproducibility within each subject, using both peak-detection consistency and FOD estimation discrepancy. For peak detection, to reduce instability caused by small spurious peaks, we restrict attention to voxels identified as having $0$, $1$, $2$, or $3$ peaks in both the test and retest scans, with the median percentage (across 37 subjects) of such voxels ranging from 59\% to 66\%, depending on the method. We compute, among these voxels, the proportion for which the same number of peaks is identified in the test and retest scans. For FOD discrepancy, we compute the median entropy-regularized OT distance between the estimated FODs of matched voxels from the test and retest scans.

\begin{figure}[ht]
    \centering
    \includegraphics[width=0.9\textwidth]{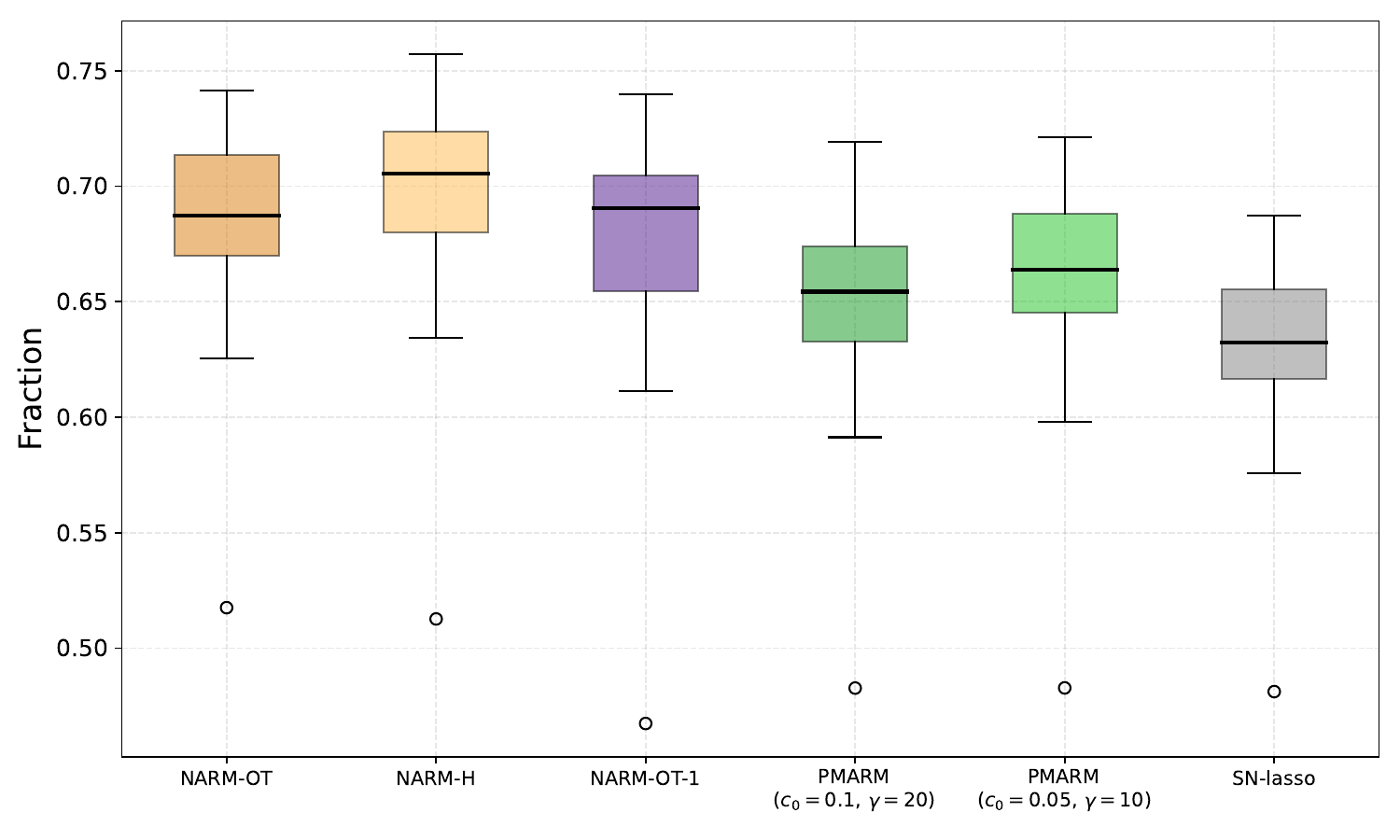}
\caption{HCP application: box plots across 37 subjects showing the fraction of voxels for which the estimated number of peaks (up to 3) agrees between test and retest scans. The fraction is computed among voxels identified as having 0, 1, 2, or 3 peaks in both scans.}\label{fig:hcp_fraction}
\end{figure}

\begin{figure}[ht]
    \centering
    \includegraphics[width=0.9\textwidth]{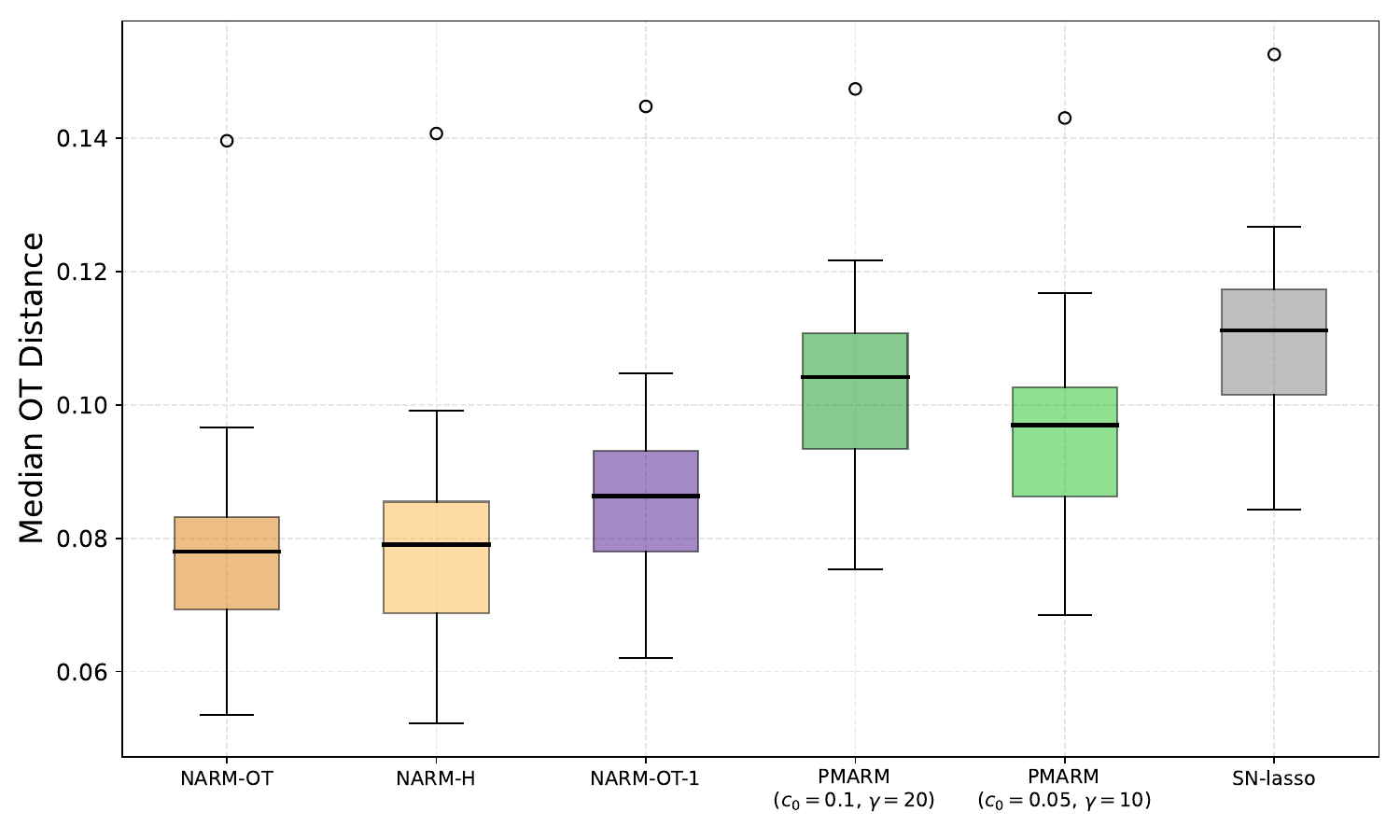}
\caption{HCP application: box plots across 37 subjects of the median entropy-regularized OT distance between estimated FODs at matched voxels across test and retest scans.}
\label{fig:hcp_OT}
\end{figure}

As summarized in Figures~\ref{fig:hcp_fraction} and \ref{fig:hcp_OT}, all three \texttt{NARM} methods outperform \texttt{PMARM} and voxel-wise \texttt{SN-lasso} in terms of peak-detection reproducibility. In terms of OT distance, the two-stage \texttt{NARM-OT} and \texttt{NARM-H} methods achieve the smallest test--retest discrepancies, followed by the one-stage \texttt{NARM-OT-1}, and then \texttt{PMARM}. All smoothing-based methods outperform voxel-wise \texttt{SN-lasso}.

%[\textcolor{red}{We further used binary masks from \textit{AutoPtx} \citep{autoptx} for streamline selection to dissect the SLF and the CST from the initial tractography results (Figure \ref{fig:streamline_selection}). [Q: did we actually do streamline selection when applying NARM?? I think we applied NARM  to a regular rectangle ROI]}]

%\begin{itemize}
%	\item Design matrix: $l_{\max} = 8$(spherical harmonic level), $j_{\max} = 3$(spherical needlet level)
%	\item $\gamma = 4$ spatial kernel parameter
%	\item $\alpha = 0.15$ weight rescaling parameter, not sensitive b/w 0.1 and 0.2.
%	\item $r$ = 1.15 (The common ratio in the radius sequence)
%	\item $S = 6$ the parameter value of neighborhood setting
%\end{itemize}

%\subsection{The execution time (hr.)}
%The execution time of NARM is measured across 70 selected ROI (test - 35, retest - 35) at each stage. S = 0 is \texttt{SN-lasso}.

%\begin{table}[ht]
%\resizebox{\columnwidth}{!}{%
%\begin{tabular}{|c|c|c|c|c|c|c|c|c|}
%\hline
%\textbf{S}    & \textbf{0}   & \textbf{1}   & \textbf{2}   & \textbf{3}   & \textbf{4}   & \textbf{5}   & \textbf{6}   & \textbf{Total} \\ \hline
%\textbf{Time} & 20.76 (5.46) & 18.57 (5.85) & 18.86 (6.17) & 16.78 (5.83) & 16.02 (5.74) & 14.68 (5.96) & 12.60 (5.60) & 118.28 (32.78) \\ \hline
%\end{tabular}%
%}
%\end{table}

%\subsection{Peak-detection algorithm parameters}
%\begin{itemize}
%	\item Number of neighborhood: 40
%	\item Threshold: 0.4 (In the region, ignore the gradient directions whose FOD is less than 0.4 of highest FOD in the neighborhood)
%	\item Maximum possible number of peaks: 4
%\end{itemize}

\section{Discussion}\label{sec:conclusions}

In this paper, we have proposed the \textit{Nearest-Neighbor Adaptive Regression Model} (\texttt{NARM}) for estimating fiber orientation distributions (FODs) from D-MRI data within a spatial ROI. By adaptively borrowing information from neighboring voxels, \texttt{NARM} incorporates spatial structure into FOD estimation and improves upon voxel-wise methods that do not use spatial smoothing. Compared with the existing spatial smoothing method \texttt{PMARM} \citep{rao2016sr}, \texttt{NARM} offers several methodological innovations: the use of optimal transport distance to better distinguish FODs with different peak configurations; a rescaling scheme that handles extreme smoothing regimes; a stopping rule based on minimum nearest-neighbor dissimilarity that is both easy to implement and effective in preventing over-smoothing; and a data-driven choice of the similarity-smoothing parameter that adapts to local fiber complexity. Simulation studies show that \texttt{NARM} yields more accurate FOD estimates than both voxel-wise \texttt{SN-lasso} and \texttt{PMARM}. When applied to HCP test--retest D-MRI data, \texttt{NARM} also produces more reproducible FOD estimates across repeated scans.

Although \texttt{NARM} was developed specifically for FOD estimation from D-MRI data, several of its methodological components are broadly applicable to spatial smoothing problems beyond this setting. In particular, the use of OT distance as a dissimilarity measure, the rescaling scheme, the voxel-adaptive stopping rule, and the data-driven selection of the smoothing parameter may prove useful in other applications involving spatially structured probability distributions or heterogeneous spatial signals.

%However, the computational efficiency is a big issue since the computation cost of the earth-mover distance is much expensive than that of the Hellinger distance.
%%%%%%%%%%%%%%%%%%%%%%%%%%%%%%%%%%%%%
%----------------- bibliography -------------------

\section*{Acknowledgement}
This research was supported by Global - Learning \& Academic research institution for Master’s·PhD students, and Postdocs (LAMP) Program of the National Research Foundation of Korea (NRF) grant funded by the Ministry of Education(RS-2024-00443714) (SYH), and by UCD Dissertation Year Fellowship (JLY), NIH 1R01EB021707 (JLY and JP), NSF-DMS-1148643 (JP) and NSF-DMS-1915894 (JP, SYH).

\bibliography{narmreference}

%%%%%%%%% appendix
%\section*{Appendix}
%\setcounter{subsection}{0}
%\renewcommand{\thesubsection}{A.\arabic{subsection}}
%\setcounter{equation}{0}
%\renewcommand{\theequation}{A.\arabic{equation}}

\phantomsection\label{supplementary-material}
\bigskip

\clearpage
\begin{center}

{\large\bf SUPPLEMENTARY MATERIAL}
\end{center}

\setcounter{page}{1}
\setcounter{section}{0}
\renewcommand{\thesection}{S.\arabic{section}}
\setcounter{subsection}{0}
\renewcommand{\thesubsection}{S.\arabic{section}.\arabic{subsection}}
\setcounter{equation}{0}
\renewcommand{\theequation}{S.\arabic{equation}}
\setcounter{figure}{0}
\renewcommand{\thefigure}{S.\arabic{figure}}
\setcounter{table}{0}
\renewcommand{\thetable}{S.\arabic{table}}
\setcounter{proposition}{0}
\renewcommand{\theproposition}{S.\arabic{proposition}}
\setcounter{lemma}{0}
\renewcommand{\thelemma}{S.\arabic{lemma}}
\setcounter{corollary}{0}
\renewcommand{\thecorollary}{S.\arabic{corollary}}

%\section{Appendix}
%For figures in the supplementary material, check out \url{https://github.com/jie108/FOD_Narm_codes/blob/master/narm_arxiv.pdf}.

\section{Entropy-regularized OT distance}
\label{sec:OT}

We consider the entropy-regularized optimal transport (OT) distance for comparing two discrete probability measures $\bm{p}, \bm{q} \in \mathbb{R}^D$ supported on a common spherical grid. Let $\bm{M}\in \mathbb{R}^{D\times D}$ denote the ground cost matrix, where $M_{ij}$ represents the cost of transporting mass from location $i$ to location $j$. In the context of FOD comparison, $\bm{M}$ is constructed using the geodesic (arc-length) distance on the real projective space $\mathbb{RP}^2$, which identifies antipodal points on the sphere and reflects the antipodal symmetry of FODs. Specifically, if $\bm{u}_i,\bm{u}_j\in \mathbb{S}^2$ are two grid directions, then
\[
M_{ij}
=
d_{\mathbb{RP}^2}(\bm{u}_i,\bm{u}_j)
:=
\arccos\!\bigl(|\bm{u}_i^\top \bm{u}_j|\bigr),
\]
which is the smaller arc-length between the two directions after antipodal identification.

The entropy-regularized distance is defined as:

\[
d^{\kappa}_{OT}(\bm{p},\bm{q}):=
\min_{\bm{\Pi} \in \mathcal{U}(\bm{p},\bm{q})}
\langle \bm{\Pi}, \bm{M} \rangle
+ \kappa \sum_{i,j} \Pi_{ij} \bigl(\log \Pi_{ij} - 1\bigr),
\]
where $\mathcal{U}(\bm{p},\bm{q}) := \{\bm{\Pi} \ge 0: \bm{\Pi}\mathbf{1}=\bm{p},\, \bm{\Pi}^\top \mathbf{1}=\bm{q}\}$ is the set of admissible transport plans. The regularization parameter $\kappa \ge 0$ controls the amount of entropy smoothing, with $\kappa = 0$ recovering the unregularized OT problem.

%\textcolor{red}{As with the Hellinger distance, we rescale this distance to lie in $[0,1]$}.

For $\kappa>0$, this problem can be solved efficiently using the Sinkhorn algorithm \citep{cuturi2013sinkhorn}, exploiting the fact that the optimal plan admits the form $\bm{\Pi} = \mathrm{diag}(\bm{u})\, \bm{K}\, \mathrm{diag}(\bm{v})$, where $\bm{K}=\exp(-\bm{M}/\kappa)$. The algorithm iteratively updates the scaling vectors $\bm{u}$ and $\bm{v}$ via matrix--vector multiplications to enforce the marginal constraints, yielding a fast and numerically stable procedure.

\section{Simulation experiments: additional tables}

The following tables report ``Co'' denotes the proportion of correctly identified voxels within each category. The notation ``Ov.$x$'' and ``Un.$x$'' denote the proportion of voxels that are over-identified or under-identified as category $x$, respectively. For example, under the 0-Fiber category, ``Ov.1'' denotes the proportion of isotropic voxels misidentified as single-fiber voxels, whereas under the 1-Fiber category, ``Un.0'' denotes the proportion of single-fiber voxels misidentified as isotropic voxels.  

\shiftedlongtable{-2.1cm}{
\begingroup
\small
\setlength{\tabcolsep}{4pt}
\renewcommand{\arraystretch}{1.1}
\begin{longtable}{c|cccc|cccc|cccc}
    \hline\hline
    & \multicolumn{4}{c|}{0-Fiber} & \multicolumn{4}{c|}{1-Fiber} & \multicolumn{4}{c}{2-Fiber} \\
    Estimator
    & Co. & Ov.1 & Ov.2 & Ov.Multi
    & Co. & Un.0 & Ov.2 & Ov.Multi
    & Co. & Un.0 & Un.1 & Ov.Multi
    \\\hline
    \texttt{NARM-OT}
    & $\bm{0.979}$ & 0.004 & 0.017 & 0
    & $\bm{1}$ & 0 & 0 & 0
    & 0.741 & 0 & 0.259 & 0
    \\\hline
    \texttt{NARM-H}
    & $\bm{0.979}$ & 0.013 & 0.008 & 0
    & $\bm{1}$ & 0 & 0 & 0
    & 0.749 & 0 & 0.251 & 0
    \\\hline
    \texttt{NARM-OT-1}
    & $\bm{0.979}$ & 0.017 & 0.004 & 0
    & $\bm{1}$ & 0 & 0 & 0
    & $\bm{0.761}$ & 0 & 0.239 & 0
    \\\hline
    \texttt{PMARM}($c_0=0.1$, $\gamma = 20$)
    & 0.950 & 0.013 & 0.038 & 0
    & 0.998 & 0 & 0.002 & 0
    & 0.700 & 0 & 0.239 & 0.061
    \\\hline
    \texttt{PMARM}($c_0=0.05$, $\gamma = 10$)
    & 0.950 & 0.013 & 0.038 & 0
    & $\bm{1}$ & 0 & 0 & 0
    & 0.697 & 0 & 0.236 & 0.067
    \\\hline
    \texttt{SN-lasso}
    & 0.950 & 0.013 & 0.038 & 0
    & 0.998 & 0 & 0.002 & 0
    & 0.685 & 0 & 0.233 & 0.082
    \\\hline\hline
    \caption{Simulation results: $(n,b,SNR) = (41, 1000, 20)$}
    \label{tab:simu-tab1}
\end{longtable}
\endgroup
}

\shiftedlongtable{-2.1cm}{
\begingroup
\small
\setlength{\tabcolsep}{4pt}
\renewcommand{\arraystretch}{1.1}
\begin{longtable}{c|cccc|cccc|cccc}
    \hline\hline
    & \multicolumn{4}{c|}{0-Fiber} & \multicolumn{4}{c|}{1-Fiber} & \multicolumn{4}{c}{2-Fiber} \\
    Estimator
    & Co. & Ov.1 & Ov.2 & Ov.Multi
    & Co. & Un.0 & Ov.2 & Ov.Multi
    & Co. & Un.0 & Un.1 & Ov.Multi
    \\\hline
    \texttt{NARM-OT}
    & $\bm{0.811}$ & 0.122 & 0.063 & 0.004
    & $\bm{1}$ & 0 & 0 & 0
    & 0.758 & 0 & 0.239 & 0.003
    \\\hline
    \texttt{NARM-H}
    & $\bm{0.811}$ & 0.126 & 0.063 & 0
    & $\bm{1}$ & 0 & 0 & 0
    & 0.758 & 0 & 0.239 & 0.003
    \\\hline
    \texttt{NARM-OT-1}
    & 0.803 & 0.088 & 0.084 & 0.025
    & $\bm{1}$ & 0 & 0 & 0
    & $\bm{0.767}$ & 0 & 0.233 & 0
    \\\hline
    \texttt{PMARM}($c_0=0.1$, $\gamma = 20$)
    & 0.454 & 0 & 0 & 0.546
    & $\bm{1}$ & 0 & 0 & 0
    & 0.706 & 0 & 0.227 & 0.067
    \\\hline
    \texttt{PMARM}($c_0=0.05$, $\gamma = 10$)
    & 0.454 & 0 & 0 & 0.546
    & $\bm{1}$ & 0 & 0 & 0
    & 0.714 & 0 & 0.233 & 0.052
    \\\hline
    \texttt{SN-lasso}
    & 0.454 & 0 & 0 & 0.546
    & $\bm{1}$ & 0 & 0 & 0
    & 0.679 & 0 & 0.227 & 0.093
    \\\hline\hline
    \caption{Simulation results: $(n,b,SNR)= (81, 3000, 10)$}
        \label{tab:simu-tab2}
\end{longtable}
\endgroup
}

\shiftedlongtable{-2.1cm}{
\begingroup
\small
\setlength{\tabcolsep}{4pt}
\renewcommand{\arraystretch}{1.1}
\begin{longtable}{c|cccc|cccc|cccc}
    \hline\hline
    & \multicolumn{4}{c|}{0-Fiber} & \multicolumn{4}{c|}{1-Fiber} & \multicolumn{4}{c}{2-Fiber} \\
    Estimator
    & Co. & Ov.1 & Ov.2 & Ov.Multi
    & Co. & Un.0 & Ov.2 & Ov.Multi
    & Co. & Un.0 & Un.1 & Ov.Multi
    \\\hline
    \texttt{NARM-OT}
    & $\bm{0.908}$ & 0.071 & 0.021 & 0
    & $\bm{1}$ & 0 & 0 & 0
    & 0.784 & 0 & 0.216 & 0
    \\\hline
    \texttt{NARM-H}
    & $\bm{0.908}$ & 0.071 & 0.021 & 0
    & $\bm{1}$ & 0 & 0 & 0
    & 0.776 & 0 & 0.224 & 0
    \\\hline
    \texttt{NARM-OT-1}
    & $\bm{0.908}$ & 0.029 & 0.029 & 0.034
    & $\bm{1}$ & 0 & 0 & 0
    & 0.784 & 0 & 0.216 & 0
    \\\hline
    \texttt{PMARM}($c_0=0.1$, $\gamma = 20$)
    & 0.513 & 0 & 0.004 & 0.483
    & $\bm{1}$ & 0 & 0 & 0
    & $\bm{0.793}$ & 0 & 0.207 & 0
    \\\hline
    \texttt{PMARM}($c_0=0.05$, $\gamma = 10$)
    & 0.513 & 0 & 0 & 0.487
    & $\bm{1}$ & 0 & 0 & 0
    & 0.787 & 0 & 0.213 & 0
    \\\hline
    \texttt{SN-lasso}
    & 0.513 & 0 & 0.004 & 0.483
    & $\bm{1}$ & 0 & 0 & 0
    & 0.787 & 0 & 0.213 & 0
    \\\hline\hline
    \caption{Simulation result:  $(n,b,SNR) = (81, 3000, 20)$ }
        \label{tab:simu-tab3}
\end{longtable}
\endgroup
}

\section{HCP application: additional details}

\subsection{Preprocessing steps}
\label{subsec:hcp_preprocessing}
The D-MRI data were downloaded from the HCP database, \textit{ConnectomeDB}, which had already undergone basic quality control and minimal preprocessing steps including intensity normalization, EPI distortion correction, Eddy current correction, Gradient nonlinearity correction, registration of the mean $b_0$ image (T2w image) to the native volume T1w image, and transformation of diffusion data, gradient deviation, and the gradient directions to the \textit{structural space (T1w space)} \citep{GLASSER2013}.
The HCP D-MRI data were additionally co-registered to the T1w space. 
We performed additional processing steps using the T1w and T2w images as described below. We used the software \textit{FSL} version 6.0.0 \citep{JENKINSON2012}, and R packages \textit{fslr} \citep{fslr} and \textit{neurohcp} \citep{neurohcp} from the \textit{neuroconductor} repository.

The (original) T1w image contained both skull and the brain;  
%Since the HCP D-MRI data were already co-registered to the structural (T1w) space, 
%so we first applied the T2w extracted binary brain mask provided by HCP to the T1w image to obtain the T1w extracted brain image. 
the T2w extracted binary brain mask provided by HCP was therefore applied to the T1w image to obtain the T1w extracted brain image.
The T1w extracted brain image and the \textit{FAST} segmentation algorithm \citep{Zhang2001} in \textit{FSL} were then used to classify each voxel into three different tissue types: CSF–cerebrospinal fluid, GM–grey matter, and WM–white matter, yielding a \textit{white-matter mask}. Voxels within the white-matter mask are hereafter referred to as \textit{white-matter voxels}.
Finally, the T1w images were registered to a standard space, \textit{MNI152-T1 2mm} (\url{http://www.bic.mni.mcgill.ca/ServicesAtlases/HomePage}), using the \textit{FSL} registration tools FLIRT \citep{JENKINSON2002825} (for initial linear registration) and FNIRT \citep{WOOLRICH2009S173} (for subsequent nonlinear registration).

\begin{figure}[ht] 	\includegraphics[width=\textwidth]{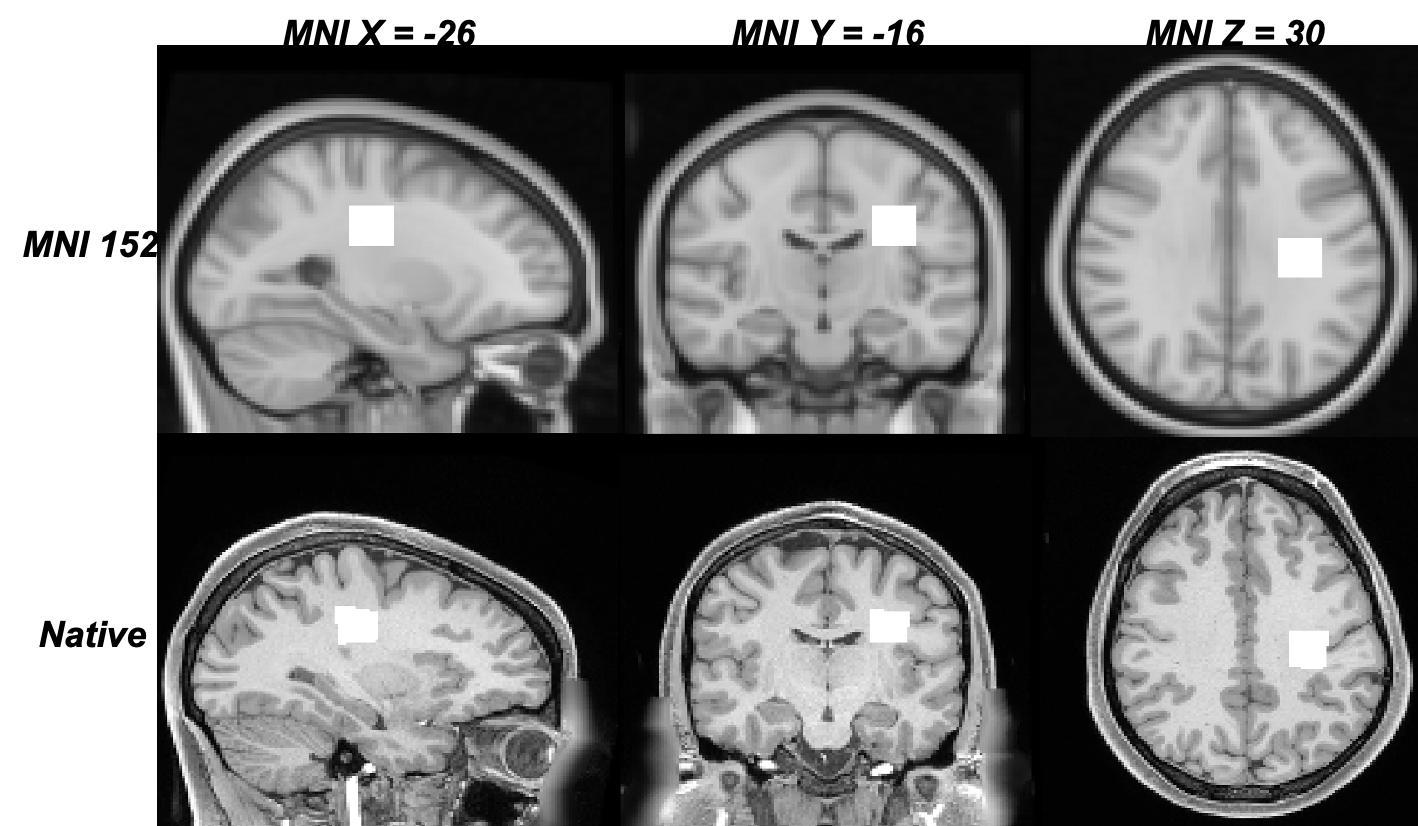}
\caption{Selected ROI used in the HCP application. Top: the ROI (white box) overlaid on the MNI152-T1 template. Bottom: the corresponding ROI (white region) mapped to the native space of a representative subject.} \label{fig:registration}
\end{figure}

\subsection{Single tensor model}\label{subsec:supp_single_tensor_model}
The single-tensor model \citep{le1995diffusion, basser2002diffusion, mori2007introduction} represents the diffusion-weighted signal along gradient direction $\bm{u}$ as
\[
S(\bm{u}) = S_0 \exp(-b\,\bm{u}^\top \bm{D}\bm{u}),
\]
where $\bm{D}$ is a $3\times 3$ positive definite matrix, referred to as the diffusion tensor. Let $\lambda_1$, $\lambda_2$, and $\lambda_3$ denote the eigenvalues of $\bm{D}$. The fractional anisotropy (FA) is defined by
\[
\mathrm{FA}
=
\frac{\sqrt{(\lambda_1-\lambda_2)^2+(\lambda_2-\lambda_3)^2+(\lambda_3-\lambda_1)^2}}
{\sqrt{2(\lambda_1^2+\lambda_2^2+\lambda_3^2)}}.
\]

The quantity $\mathrm{FA}\in[0,1]$ measures the degree of diffusion anisotropy: larger FA indicates stronger directional diffusion, typically along the principal eigenvector of $\bm{D}$, whereas smaller FA suggests more isotropic diffusion, provided that the single-tensor model is adequate. However, in the presence of multiple crossing fibers, FA can also be small, which may incorrectly suggest isotropic diffusion. This inability to resolve crossing fibers is a major limitation of the single-tensor model and motivates the development of more flexible models, including the FOD model.
%In contrast, MD measures the overall magnitude of diffusion, with larger MD corresponding to faster diffusion and smaller MD to slower diffusion.

\subsection{Response function estimation}
\label{subsec:hcp_response}

We estimate the response function $R(\cdot)$ in the FOD model \eqref{eq:spherical_convolution_model} separately for each D-MRI scan, following the procedure of \citet{yan2018estimating}. First, we fit the single-tensor model to every white-matter voxel in the brain. Voxels with fractional anisotropy (FA) greater than $0.8$ and a ratio of the larger to the smaller of the two minor eigenvalues of the estimated diffusion tensor less than $1.5$ are classified as having a single dominant fiber bundle.

Let $\bar{\lambda}$ and $\underline{\lambda}$ denote the medians of the largest eigenvalue and the smaller eigenvalues, respectively, across these voxels. The response function is then specified as
\[
R(\cos\theta)
:=
S_0 \exp\!\left\{-b\left(\bar{\lambda}\cos^2\theta+\underline{\lambda}\sin^2\theta\right)\right\},
\qquad \theta\in[0,\pi].
\]
Finally, we normalize the DWI measurements at each voxel by the mean intensity of the six $b_0$ images at that voxel and set $S_0=1$ in the response function.

\end{document}